\documentclass[12pt,a4paper]{article}
\usepackage{graphicx}
\usepackage[cp1251]{inputenc}
\usepackage{amssymb, amsfonts, amsmath}
\usepackage{amscd}
\begin{document}

\title{Separation of variables in Hamilton-Jacobi equation for a charged test particle in the Stackel spaces of type (2.0)}

\author{V. V. Obukhov}

\date{}

\maketitle
\abstract Found all equivalence classes for electromagnetic potentials and space-time metrics of Stackel spaces, provided that the equations of motion of the classical charged test particles are integrated by the method of complete separation of variables in the Hamilton-Jacobi equation. Separation is carried out using the complete sets of reciprocal-commuting integrals of motion of type (2.0), whereby in a privileged coordinate system the Hamilton-Jacobi equation turns into a nonparabolic type equation.

\quad


\quad
\begin{flushleft}

           \begin{center}
            Tomsk State  Pedagogical University, 60 Kievskaya St., Tomsk, 634041, Russia.

            Tomsk State University of Control Systems and Radioelectronics 40 Lenin Ave., Tomsk, 634050, Russia
           \end{center}
\end{flushleft}

\section{Introduction}

The construction of the theory of the separation of variables in the classical and quantum equations of test particles motion, as well as in the classical equations of mathematical physics, is one of the most important results in the theory of symmetry implementation. The main tool in the theory of the separation of variables is the Stackel spaces. Thus, the Riemannian spaces, whose metrics in some (privileged) coordinate systems allow separation of variables in the free Hamilton-Jacobi equation were named after Paul Stackel  \cite{1}, \cite{2}.
In the found metric, the equation admits a complete separation of variables in the orthogonal privileged coordinate system. The equation itself has a complete set of motion integrals quadratic in velocity. Among the articles devoted to the Stackel spaces beside the works from Stackel  himself and from Eisenhart \cite{3}, we mention Levy-Civita \cite{4}, who formulated in a covariant form the condition for separation of variables in privileged coordinate systems, and Yarov-Yarovoy \cite{5}, who generalized the definition of Stackel spaces in the case of non-orthogonal privileged coordinate systems. V.N. Shapovalov in a series of works \cite{6} - \cite{8} proved the main theorem of the theory of separation of variables - the theorem on the necessary and sufficient conditions for separation. In the case of the free Hamilton-Jacobi equation, such condition is met when there is a complete set of Killing fields in space. The set includes $n$ geometric objects which are called Killing tensor fields not higher than the second rank, which mutually commute and meet some additional restrictions.

The results obtained by solving the appropriate Killing equations allowed finding all Stackel metrics in the real coordinate systems. The work \cite{9} provides proof for the generalization of the theorem in the case of complex privileged coordinate systems.

 Even though there are known examples of spaces in which the quantum-mechanical equations of motion allow non-commutative integration (see, for example, \cite{10}), these were Stackel spaces that attracted the greatest interest of researchers due to their high level of symmetry and rich geometric content.  So, the types of Stackel metrics for the flat space-time were used to construct the theory of separation of variables in the Klein-Gordon-Fock and Dirac equations.  Using the complete sets of symmetry operators, all privileged coordinate systems, electromagnetic potentials, and many examples of the exact integration of equations by the method of complete separation of variables were found (see, for example: \cite{11}-\cite{14} ).

A special place is taken by the Stackel spaces in the theory of gravitation. The set of Stackel spaces of type $(2.0)$ include such important solutions of the Einstein equations as solutions found by Schwarzschild, Kerr, and by   Taub-Newman-Unti-Tambourino (Taub-NUT solution). Moreover, the basic metric of cosmology, the Roberson-Walker metric also belongs to the set of  Stackel spaces, namely, the spaces of type $(1.0)$. A high level of symmetry of the metric allows to provide new approaches to solving cosmological problems, in particular, related to the consideration of dark energy or to the construction of modified gravity theories (see, for example, \cite{15}, \cite{16}). Also, it is possible to use methods of symmetry theory to justify  model selection in the Extended Gravity Cosmology \cite{17}.

Hundreds of works have been devoted to the study of the Stackel spaces geometry  so as  the problem of integrating the basic equations of mathematical physics and field equations of the theory of gravitation. A sufficiently detailed bibliography, can be supplemented by the numerous works of Russian authors, can be found in papers \cite{18} - \cite{18a}.

Recently, interest in the problem of separation of variables has grown significantly. So the work \cite{19}  separation of variables of type (2.0) is directed to obtain the exact solutions of the Maxwell equation in the Plebanski-deminski space. A lot of work is devoted to the study of various effects in individual Stackel spaces; see, for example, \cite{21}-\cite{30a}.

 One of the conclusions of the necessary and sufficient conditions theorem is the partition of the set of Stackel metrics into invariant subsets according to the type of the complete set of Killing fields.
According to the definition of the metric of Stackel spaces, the complete sets and the complete separation of variables are of type $(N.N_0)$,
 where $N$ is the maximum number of independent Killing vector fields ${Y_p}^i$, included in the complete set; $ N_0 = N-\det({Y_p}^i{Y_q}_i) $.  In the case of the Lorentz signature $N_0=0,1$.  If $N_0=0$, the space is called non-null. Otherwise, null.
 There are seven disjoint sets of Stackel spaces with the signature of the space-time.

1) Non-null Stackel spaces of type (3.0). The complete set includes three Killing vector fields. The coordinate hypersurface related to the non-ignored variable is non-null. (Ignored are the privileged variables (see the definition below) that occur linearly in the complete integral. Other variables are called non-ignored.)

2) Null Stackel spaces of type (3.1).  The complete set includes three Killing vector fields.  The coordinate hypersurface related to the non-ignored variable is null one.

3) Non-null Stackel spaces of type (2.0).  The complete set includes two Killing vector fields. Coordinate hypersurfaces related to non-ignored variables are non-null.

4) Null Stackel spaces of type (2.1).  The complete set includes two Killing vector fields.  The coordinate hypersurface belonging to one of the non-ignored variables is null.

5) Non-isotropic Stackel spaces of type (1.0). The complete set includes one non-isotropic Killing vector field. Coordinate hypersurfaces related to non-ignored variables are non-null.

6) Null Stackel spaces of type (1.1). The set includes one isotropic Killing vector field. The coordinate hypersurface belonging to one of the non-ignored variables is null.

7) Non-null Stackel spaces of type (0.0). The complete set does not contain Killing vector fields. Coordinate hypersurfaces related to non-ignored variables are non-null.

The theorem about necessary and sufficient conditions for the separation of variables made it possible to develop a theory of complete separation of variables for single-particle classical and quantum-mechanical equations of motion in the theory of gravity in the presence of physical fields of various nature. A task occurred to systematically classify Stackel spaces in the presence of these fields. By classification, we mean an enumeration of all nonequivalent relatively admissible (i.e. non-violating conditions for the complete separation of variables) coordinate transformations and gauge transformations of the potentials. Since the set of Stackel spaces is divided into disjoint subsets consisting of spaces of  type ${(N.N_0)}$, the classification is carried out separately for each type.

  Apparently, the first example of a systematic classification was made by J. Iwata \cite{31}. In the paper vacuum solutions for the Einstein equations for spaces of type $(1.0)$ were presented. In works \cite{32}-\cite{35} a similar classification for the remaining types of vacuum Stackel spaces was carried out. For electro-vacuum Stackel spaces, the classification was made in papers \cite{36}-\cite{38}.

 However, to date, the classical classification problem for the case when the Hamilton-Jacobi equation for a charged test particle moving in an external electromagnetic field in the absence of additional restrictions for this field admits complete separation of variables, has not been solved. In this work, the classification is constructed for the type \quad $(2.0)$.\quad All appropriate non-equivalent sets of space-time metrics and potentials of the external electromagnetic field are found.

integratedintegrated

  \section{Conditions for the existence of a complete set of motion integrals.}

 Let us consider spaces of type (2.0) in which the Hamilton-Jacobi equation for a charged test particle

\begin{equation}\label{1} g^{ij}P_iP_j  =  \tilde{\lambda}; \quad P_i = p_i + A_i = S_{,i}+A_i, \end{equation}

can be integrated by the complete separation of variables method. In the privileged coordinate set the complete integral
has the additive form:

\begin{equation}\label{2} S=\sum{s_{i}(u^i)}.\end{equation}

Let us recall that according to the definition the coordinate system, in which such form exists, is called privileged and denoted by variables: \quad
$u^i=(u^0,u^1,u^2,u^3)$.\quad Throughout the text, the repeating upper and lower indices are summed up within the established limits for indices changes;\quad $i,j= 0,1,2,3$.\quad
  We consider the field external when the electromagnetic potential contains at least one (free) function, which is not expressed through the metric tensor. If the complete set contains \quad $N$ \quad Killing vector fields, the first \quad $N$ \quad coordinates are ignored.  We will denote their coordinate indices by the letters\quad $ p,q, = 0,1, \dots = N-1. $ \quad We will supply the non-ignored coordinates with the indices\quad $\mu,\nu = N, \dots,3$.\quad  Functions that depend only on the variable\quad $u^2$\quad will be denoted by the lowercase Greek letters, only from the variable \quad $u^3$ \quad by the e lowercase Latin letters. Lowercase Latin and Greek letters with a tilde icon indicate constants.  Exceptions:
 $\delta^{ij}, \delta_{ij}, \delta^{i}_{j}$ are the Kronecker symbols, $g^{ij},\quad g_{ij}$ - components of the metric tensor, $\varepsilon, \varepsilon_i=+1,-1; \quad , \quad \lambda,\lambda_{i} = const $, \quad  $h_{\nu}^{ij}, h_\nu^i, h_\nu$ - functions of $u^{\nu}$.
       In this notation, the metric tensor of the Stackel space of type (2.0) can be written as:

       \begin{align}\label{3} g^{ij}= (\hat{\Phi}^{-1})_3^\nu h^{ij}_\nu =\frac{\delta^i_p\delta^j_q ( a^{pq} + \alpha^{pq})+\varepsilon^\nu\delta^i_\nu\delta^j_\nu}{\Delta}, \end{align}
  \begin{align}\label{4}\hat{(\Phi})^{\mu}_{\nu}=\begin{pmatrix} 1 & \phi\\ -1 & f \end{pmatrix},\quad \Delta = \phi+ f=\det\hat{\Phi},\end{align}
   where $ p,q, = 0,1$; $\nu, \mu = 2,3$. $\quad \hat{\Phi}$  - Stackel matrix.
   Note that the Stackel space of type $(2.0)$ admits two Killing vector fields:  $ \quad Y^i_p=\delta^i_p \quad $ and two Killing tensor vector fields (together with the metric tensor):
   \begin{align}\label{5}\quad X^{ij}_2=\frac{f\varepsilon^2(\delta^i_p\delta^j_q\alpha^{pq}+\delta^i_2\delta^j_2) -\phi\varepsilon^3(\delta^i_p\delta^j_qa^{pq} + \delta^i_3\delta^j_3)}{\Delta}. \quad  X^{ij}_3= g^{ij}.\end{align}
The complete set of integrals of motion for the free Hamilton-Jacobi equation has the form:

\begin{equation}\label{6} \hat{X_\nu} = X^{ij}_\nu p_ip_j,\quad \hat{X_q} = Y_q^ip_i. \end{equation}

The integrals of motion of the Hamilton-Jacobi equation, quadratic in impuls, included in the full set in the presence of an external electromagnetic field have the form:
\begin{equation}\label{6} \hat{X_2}=X^{ij}_2p_ip_j +2X^ip_i +X, \quad  \hat{X_3}=g^{ij}P_iP_j. \end{equation}
The complete integral in the privileged coordinate system can be reduced to the form:
\begin{align}\label{7}S=\sum_{q=0}^1\lambda_q u^p +\sum_{\nu=2}^3s_\nu.\end{align}
The presence of a complete set of commuting motion integrals allows us to find \eqref{7} as a solution to the system of equations:
\begin{align}\label{8}\hat{X}_i=\lambda_i.\end{align}
Moreover, $\hat{X}_i$ have the form:
 \begin{align}\label{9} \hat{X}_q = \delta^i_q p^q =p_q,\quad \hat{X}_\nu =(\hat{\phi}^{-1})_{\nu}^{\mu}\hat{H}_\mu = (\hat{\phi}^{-1})_{\nu}^{\mu}(h^{ij}_{\mu}p_{i}p_{i}+2h^{i}_{\mu}p_{j}+h_{\mu}),\end{align}
 where
 $$h_2^{ij} =\delta^i_p{\delta^j_q}\alpha^{pq}+\varepsilon^2\delta^i_2\delta^j_2,\quad h_2^i=\omega^i,\quad h_2=\hat{\omega},$$
 $$h_3^{ij}=\delta^i_p{\delta^j_q}a^{pq}+\varepsilon^3\delta^i_3\delta^j_3,\quad h_3^i=h^i,\quad h_3=\hat{h}.$$
Equating the coefficients before the impulses to the right and left in the  \eqref{9} when $\nu=3$ and vanishing with a gradient transformation of the function potential $\omega^2$, $h^3$, we obtain:

\begin{align}\label{10}A^p=\frac{\omega^p+h^p}{\Delta}, \quad A^\nu = 0, \quad A_{i}A^i=\frac{\hat{\omega}+\hat{h}}{\Delta}.\end{align}
The expressions \eqref{10} determine the electromagnetic potential of $A^i$.  A necessary and sufficient condition for the equation \eqref{1} to admit complete separation of variables is:
 \begin{align}\label{11}A^iA_i=\frac{(\omega^p+h^p)(\omega^q+h^q)g_{pq}}{\Delta}=(\hat{\omega}+\hat{h}).\end{align}
By setting:
\begin{align}\label{12}G^{pq}=g^{pq}\Delta, \quad G_{pq}=\frac{g_{pq}}{\Delta}, \quad  G=\det{(G^{pq})},\quad G^{pq}=\alpha^{pq}+a^{pq},\end{align}
 we obtain the necessary and sufficient condition in the form of a functional equation:
\begin{align}\label{13}(\alpha^{00}+a^{00})(\omega^{1}+h^{1})^2+(\alpha^{11}+a^{11})(\omega^{0}+h^{0})^2-&
\cr2(\alpha^{01}+a^{01})(\omega^{1}+h^{1})(\omega^{0}+h^{0})=&
\cr(det(\alpha^{pq})+det(a^{pq})+\alpha^{11}a^{00}+\alpha^{00}a^{11}-2\alpha^{01}a^{01})(\hat{\omega}+\hat{h}).\end{align}

Note that in the general case of an  $n$-dimensional Stackel space, it follows from the Shapovalovs theorem that the equation \eqref {1} admits complete separation of variables if and only if there exists a coordinate system in which:

$$  g^{ij}= (\hat{\Phi}^{-1})_{n}^\nu  h^{ij}_\nu, \quad  A^i=(\hat{\Phi}^{-1})_{n}^\nu h^{i}_\nu,$$
\begin{align}\label{14}\quad A^iA_i = (\hat{\Phi}^{-1})_{n}^\nu h_\nu.\end{align}
 $\nu =1,2,.\dots n, \quad h^{ij}_\nu,\quad h^{i}_\nu,\quad h^{(0)}_\nu,\quad  h_\nu, \quad$ - functions only from $\quad u^\nu$, $\quad \hat{\Phi}$ -Stackel matrix. If we introduce the additively arbitrary scalar field $\Psi$ into the equation \eqref {1}:

 $$ g^{ij}P_iP_j + \Psi =  \tilde{m},$$
(see \cite{39}, where the classification problem for the natural Hamilton-Jacobi equation was considered), it is easy to ensure that the necessary and sufficient conditions are satisfied by a simple choice:

 $$\Psi = (\hat{\Phi}^{-1})_{n}^\nu h^{(0)}_\nu -A^iA_i.$$
In our case, for the classification, it is necessary to solve a non-trivial functional equation \eqref{14}.
It can be shown that the solution for the functional equation  \eqref{13} is equivalent to the solution of two algebraic equation systems, each containing 7 equations (6 of the second degree and one of the third degree). The first system includes functions only from the variable $u^2$, the second only from $u^3$. Both systems are overcrowded and solutions are possible if there is an additional symmetry in the equation \eqref{13}. As a similar symmetry, justified from a physical point of view, we require that the electromagnetic field be external.  Since the variables $u^2$ and $u^3$ appear in \eqref{13} symmetrically, without loss of generality, we can assume that at least one free function is contained in $\omega^p$. We also assume that all nonzero functions in \eqref{13} are fairly smooth, and there are points on the coordinate axes in the vicinity of which not one of them vanishes.
\maketitle

\section{Solutions for the case when both functions $\omega^p$ are free.}
   To solve the classification problem, it is necessary to find and resolve the conditions for the existence of the free function. First, consider the case where both $\omega^p$ are free. We use the smoothness condition mentioned above and consider the equation \eqref{13} at a fixed point $\tilde{u}^3$ on the coordinate axis ($u^3$). Since $G \ne 0 ,\quad \omega \quad $ is expressed in terms of the function $\omega^{p}$ as follows:
  \begin{align}\label{15}\hat{\omega}=\gamma_{pq}\omega^{p}\omega^{q}+2\gamma_{p}\omega^{p}+\gamma.\end{align}
    Here the functions $\gamma_{pq},\gamma_{p},\gamma$ are rational functions of $\alpha^{pq}$.
   We substitute \eqref{15} into \eqref{13} and equate the coefficients of $\omega^{p}\omega^{q}$ and $\omega^{p}$ on the right and left sides in \eqref{13}. Hence, we obtain the conditions for the existence of an external electromagnetic field.
   From the equality to zero of the coefficients before $\omega^p\omega^q$ it follows:
   \begin{align}\label{16}(\alpha^{pq}+a^{pq})=\gamma_{pq}G,\quad G=\frac{1}{det(\gamma_{pq})}, \quad \to  a^{pq}=0 \to  det(\alpha^{pq})\ne 0.\end{align}
    The equality of coefficients before $\omega^p$ to zero gives:
\begin{equation}\label{17}
   \left\{\begin{array}{rl}
      \alpha^{00}h^{1}-\alpha^{01}h^{0}=\gamma_{1},\\
   \cr \alpha^{01}h^{1}-\alpha^{11}h^{0}=-\gamma_{0}.\\
     \end{array}\right.\end{equation}
 As \quad $det(\alpha_{ pq}) \ne 0, \quad  h^p=\tilde{h}^p=0 \quad \to  A^p=\frac{\omega^p}{\Delta},$\quad
  and the complete set integrals of motion has the form:
\begin{align}\label{18}\hat{X}_q=p_q, \quad \hat{X}_2=\frac{f\varepsilon^2(\alpha^{pq}p_pp_q+2\omega^{q}p_q+\alpha_{pq}\omega^p \omega^q + p_2^2) -\phi\varepsilon^3p_3^2}{\Delta}.\end{align}
\begin{align}\label{19}\hat{X}_3=g^{ij}{p_ip_j},\quad g^{ij}=\frac{\delta^i_p\delta^j_q\alpha^{pq}+\varepsilon^\nu\delta^i_\nu\delta^j_\nu}{\Delta}.\end{align}

\section{The linear dependence of $\omega^p$ on a free function.}

        Now let $\omega^0$ and $\omega^1$ be connected to each other by a linear relation of the form:
    \begin{align}\label{20}\omega^p=\alpha^p\sigma+\sigma^p.\end{align}
       Here $\sigma$ is a free function; $ \sigma^p$, $\alpha^p$ are some rational functions of $\alpha^{pq}$. Obviously, the function $\hat{\omega}$ in the relation to \eqref {13} depends on the free function as follows:
     \begin{align}\label{21}\hat{\omega}=\gamma\sigma^2+2\gamma_0\sigma+\omega_0.\end{align}
      $\gamma,\quad \gamma_0,\quad \omega_0$\quad are also rational functions of $\alpha^{pq}$.
We substitute the expressions \eqref{20} and \eqref{21} in \eqref{13}.
The equality to zero of the coefficients before \quad  $\sigma^b \quad (b=0,1,2)$ \quad  gives the system of equations:
 \begin{align}\label{22}a^{00}(\alpha^1)^2 +a^{11}(\alpha^0)^2 - 2a^{01}\alpha^0\alpha^1 +&
 \cr\alpha^{11}(\alpha^0)^2 +\alpha^{00}(\alpha^1)^2- 2\alpha^{01}\alpha^0\alpha^1=G\gamma,.\end{align}
 $$(\gamma=0,1)$$
 \begin{align}\label{23}(a^{00}+\alpha^{00})\alpha^1(h^1+\sigma^1)+(a^{11}+\alpha^{11})\alpha^0(h^0+\sigma^0) & -\cr(a^{01}+\alpha^{01})(\alpha^0(h^1+\sigma^1)+\alpha^1(h^0+\sigma^0))=G\gamma_0 ,\end{align}
 \begin{align}\label{24}(\alpha^{00}+a^{00})(\sigma^{1}+h^{1})^2+(\alpha^{11}+a^{11})(\sigma^{0}+h^{0})^2-&
\cr2(\alpha^{01}+a^{01})(\sigma^{1}+h^{1})(\sigma^{0}+h^{0})=
G(\omega_0+h).\end{align}
The equation \eqref{22} is a condition for the existence of a free function, which is superimposed on the metric tensor. Spaces of type (2.0) satisfying \eqref{22} already allow complete separation of variables in the equation \eqref{1}. The remaining equations of the system serve to find the functions \quad $h^p$.\quad Note that they have the obvious solution \quad $h^p=0$.\quad
To find another solutions of the equation \eqref{22}, we classify the matrices \quad$\hat{G}= (G^{pq})$,\quad with respect to the group of admissible transformations of ignored variables: \quad$u^p\rightarrow \tilde{c}^{p}_qu^q$.\quad Let us list all equivalence classes of matrices $ \hat{G}$. The classes will be denote: \quad $\hat{G}_\alpha$,\quad where\quad $\alpha = 1, \dots 5$.

   \begin{equation}\label{25}
   \left\{\begin{array}{rl} \displaystyle \cr\vspace{3mm}
&     \hat{G}_1=\displaystyle\begin{pmatrix} a_0+\alpha^{00} & a+\alpha^{01} \\ a+\alpha^{01} & a_1+\alpha^{11} \end{pmatrix},\quad \cr\vspace{5mm}
  & \hat{G}_2=\displaystyle\begin{pmatrix} a_0+\alpha^{00} & a+\alpha^{01} \\ a+\alpha^{01}& -a_0+\alpha^{11} \end{pmatrix}, \cr\vspace{3mm}
   &\hat{G}_3=\displaystyle\begin{pmatrix} a_0+\alpha^{00} & \alpha^{01} \\ \alpha^{01} & a_1+\alpha^{11} \end{pmatrix},\quad \cr\vspace{3mm}
   &\hat{G}_4=\displaystyle\begin{pmatrix} a_0+\alpha^{00} & a+\alpha^{01} \\ a+\alpha^{01} & \alpha^{11} \end{pmatrix},\quad \cr\vspace{3mm}
   & \hat{G}_5=\begin{pmatrix} a+\alpha^{00} & \alpha^{01} \\ \alpha^{01} & \varepsilon a+\alpha^{11} \end{pmatrix}, \quad \varepsilon = 0,-1.\\
     \end{array}\right.\end{equation}
  Here \quad $a^{00}= a_0,\quad a^{01}=a,\quad a^{11}=a_1$.\quad Functions \quad $a_0,\quad a,\quad a_1 $ \quad  are linearly independent.

\subsection{Finding the metric tensor.}
  As already noted, the equation \eqref{22} does not contain functions defining the electromagnetic potential and, being a necessary and sufficient condition for the existence of a free function, allows us to find the metric tensor. Let us present \eqref{22} in the following form:
 \begin{align}\label{26}(a_0{\alpha^1}^2 +a_1{\alpha^0}^2-2a\alpha^0\alpha^1) + (\alpha^{11}{\alpha^0}^2 +\alpha^{00}{\alpha^1}^2-  2\alpha^{01}\alpha^0\alpha^1)=& \cr(det(\alpha^{pq})+det(a^{pq})+\alpha^{11}a_0+\alpha^{00}a_1-2\alpha^{01}a) \tilde{\gamma},\end{align}
 and consider it separately for options: $\tilde{\gamma}=1$ and $\tilde{\gamma}=0$.

      $\mathbf{I}.$ $\tilde{\gamma}=1.$
      We show that in this case all the solutions of the equation \eqref{22} for all classes $\hat{G}_\alpha$ from the expressions \eqref{25} except $\hat{G}_5$ can be represented as:
        \begin{align}\label{27}
  \hat{G}=\begin{pmatrix}(a^0)^2+\varepsilon(\alpha^0)^2 & a^0a^1+\varepsilon\alpha^0\alpha^1 \\ a^0a^1+\varepsilon\alpha^0\alpha^1 & (a^1)^2+\varepsilon(\alpha^1)^2 \end{pmatrix}.\end{align}
   By marking: $ \quad\beta^{pq}=\alpha^p\alpha^q-\alpha^{pq}, \quad $   we bring the equation \eqref{22} to the form:
  \begin{align}\label{28}\det(a^{pq})+det(\beta^{pq})+a_1\beta^{00} +a_0\beta^{11}-2a\beta^{01}=0.\end{align}
 To prove the statement, let us consider \eqref{28} for all $\hat{G}_\alpha$  from the expressions \eqref{25} with numbers $ \alpha =1, \dots 4 $ .

 $\mathbf{a)}.$ $\hat{G}=\hat{G}_1$.

   Let us fix the variable $u^{2}$ to the point ${ \quad u^2=\tilde{u}^2} \quad $ in the functional equation \eqref{23}. As result, we get the
expression:
    $$det(a^{pq})=\tilde{\gamma}_{pq}a^{pq}+\tilde{a},$$ which after the shift of the functions $\alpha^{pq} \to \alpha^{pq}+\gamma_{pq}$, and vanishing $\tilde{\gamma}_{pq}$ can be present in the form:
    $ \quad det(a^{pq})=\tilde{a}.\quad $ Since all $a^{pq}$ are linearly independent, from \eqref{28} it follows:
    $$\beta^{pq}=0 \rightarrow \alpha^{pq}=\alpha^p\alpha^q,\quad \tilde{a}=det(a^{pq})=0.$$
From $\quad deta^(pq)0 \quad $ we get: $ \quad a^{pq}=a^pa^q.\quad $ Therefore, the matrix $\quad \hat{G} \quad$ can be represented as \eqref{27}.  The statement is proved.

  $\mathbf{b})$ $\hat{G}=\hat{G}_2$.
The functional equation follows from \eqref{28}:
\begin{align}\label{29}\det(a^{pq})+det(\beta^{pq})+a_{0}(\beta^{11}-\beta^{00})-2a\beta^{01}=0.\end{align}

  Since in $a_{0}$ and $a$ are linearly independent, from \eqref{28} it follows:

  $$a^2+a_0^2=\tilde{a}^2, \quad \beta^{pp}=\tilde{a},\quad \beta^{01}=0.$$


 The elements of the matrix $\hat{G}$ can be represented as:
 $$a^{00}=\tilde{a}+a_{0},\quad a^{11}=\tilde{a}-a_{0},\quad a^{01}=a,\quad \alpha^{pq}=\alpha^{p}\alpha^{q}.$$
 Thus, \quad $det(a^{pq})=det(\alpha^{pq})=0$, \quad The equality to zero of the coefficients before \quad $ (\sigma)^b \quad ( b = 0,1,2 )$ \quad gives the system of equations:  and the matrix \quad $\hat{G}_2$ \quad are reduced to the form \eqref{26}.

  $\mathbf{c})$ $\hat{G}=\hat{G}_3$. The functional equation follows from \eqref{28}:
   $$a_0a_1+ \det{\beta^{pq}}+a_0 \beta^{11}+a \beta^{00} = 0.$$
  Since $\quad aa_0 \ne 0 \quad $, we get: $ \quad \beta^{pq}=\tilde{\beta}^{pq}, \quad (a_0 + \tilde{\beta}^{00})(a_1 + \tilde{\beta}^{11}) = \tilde{a}^2,\quad $
 Therefore, the matrix $\quad \hat{G}\quad $ can be represented as:
 \begin{align}\label{33}
 \hat{G}= \begin{pmatrix} a+\varepsilon{\alpha^0}^2 & \tilde{a}+\alpha^0 \alpha^1 \\ \tilde{a}+\alpha^0 \alpha^1 & (\frac{\tilde{a}^2}{a}+\varepsilon{\alpha^0}^2)\end{pmatrix}.\end{align}
 and thus, result in the form \eqref{26}.

 $\mathbf{d})$ $\hat{G}=\hat{G}_4$. The functional equation follows from \eqref{28}:
  $$a_0\beta^{11}-2a\beta^{01}+det{\beta^{pq}}=\varepsilon a^2.$$
  Since $\quad a,a_0 \ne const\quad $ we get: $\quad a_0=\varepsilon a^2,\quad\alpha^{00}=(\alpha^{0})^2,\quad\alpha^{01}=\alpha^0\alpha^1,\quad\alpha^{11}=1+(\alpha^1)^2,$ and the matrix $\hat{G}$ can be represented as:
  \begin{align}\label{34}
  \hat{G}=\begin{pmatrix}\varepsilon a^2+(\alpha^0)^2 & a+\alpha^0\alpha^1\\a+\alpha^0\alpha^1 & \varepsilon +(\alpha^1)^2\end{pmatrix}.\end{align}
  Thus, it has the form \eqref{26}.

  $\mathbf{e)}$ Now consider the case when $\hat{G}=\hat{G}_5$. The function $\quad a \ne const, \to \varepsilon=0 $,\quad and the functional equation follows from \eqref{28}:
 $$a(\alpha^{11}-{\alpha^1}^2)=(\alpha^{00}-{\alpha^0}^2)(\alpha^{11}-{\alpha^1}^2)-(\alpha^{01}-\alpha^0\alpha^1)^2.$$
 From here:  $\quad \alpha^{11}={\alpha^1}^2,\quad \alpha^{01}=\alpha^0\alpha^1,\quad $
 and the matrix  $\hat{G}$ takes the form:
 \begin{align}\label{35}
 \hat{G}=\begin{pmatrix}a+\alpha+{\alpha^{0}}^2 & \alpha^0\alpha^1\\ \alpha^0\alpha^1 & {\alpha^1}^2\end{pmatrix}.\end{align}

   $\mathbf{II}.$ $\gamma=0.$

   In case when the matrices $\hat{G}_\rho$ belong to the first three variants ($(\rho=1, \dots 3)$), the equation \eqref{22} has no nonzero solutions and $\omega^p$ does not contain any free function.
  Consider the remaining options.

  $\mathbf{a})$. $\hat{G}=\hat{G}_4$.  From the equation \eqref{22} it follows:
       \begin{align}\label{36}a_{0}{\alpha^1}^2-2a\alpha^0\alpha^1+\alpha^{11}{\alpha^0}^2 +\alpha^{00}{\alpha^1}^2- 2\alpha^{01}\alpha^0\alpha^1=0.\end{align}
        From here:$ \quad \alpha^1=\alpha^{11}=0 \quad$, and we get the matrix $ \quad \hat{G} \quad$ in the form:
        \begin{equation}\label{37}\hat{G}=\begin{pmatrix} a_0+\alpha_0 & a+\alpha \\ a+\alpha & 0 \end{pmatrix}. \end{equation}

 $\mathbf{b})$ $\hat{G}=\hat{G}_5$. From the equation \eqref{22} it follows:
 \begin{align}\label{38}{\alpha^1}^2 +\varepsilon{\alpha^0}^2=0,\quad \alpha^{11}{\alpha^0}^2 +\alpha^{00}{\alpha^1}^2- 2\alpha^{01}\alpha^0\alpha^1=0.\end{align}
       First, let $\quad \varepsilon = -1. \to \alpha^{0}=\alpha^1,\quad \alpha^{11}=-\alpha^{00} +2\alpha^{01},\quad  $
     and the matrix $\hat{G}$ takes the form:
\begin{align}\hat{G}=\begin{pmatrix}(a_{0}+\alpha_{0}) & \alpha \\ \alpha & -(a_0+\alpha_{0})+\alpha\end{pmatrix}.\end{align}
By replacing the variables:\quad
$\hat{u^0} \to \frac{1}{\sqrt{2}}(u^0+u^1),\quad \hat{u^1} \to \frac{1}{\sqrt{2}}(u^0-u^1)$,
we bring the solution to the form:
\begin{align}
\hat{G}=\begin{pmatrix}\alpha & (a_0+\alpha_{0}) \\(a_0+\alpha_{0}) & 0\end{pmatrix}.  \end{align}
Now let  $\varepsilon=0$. It is easy to show that in this case  $\alpha^1 = 0$ and $\hat{G}$ has the form:
\begin{equation}\label{39}\hat{G}=\begin{pmatrix} a_0+\alpha_0 & \alpha \\ \alpha & 0 \end{pmatrix}, \quad \alpha^{1}=0. \end{equation}
Thus, both solutions are special cases of the solution \eqref{37}.

\subsection{Building an electromagnetic potential.}

 To complete the classification, it is necessary to establish the dependence of the electromagnetic potential on the variable $u^3$ for this, it is necessary to consider the remaining equations \eqref{23} and \eqref{24}.

$\mathbf{a})$ Matrix $\hat{G_1}$.
We substitute the matrix $\hat{G_1}$ into the equation \eqref{23}. After some transformation  we get the equation:
$$a^0h^1 -a^1h^0=a^0(\gamma\alpha^1-\sigma^1)-a^1(\gamma\alpha^0-\sigma^0).$$
This implies:
$$a^0(h^1-\tilde{c}^1)=a^0(h^0-\tilde{c}^0),\quad \sigma^p=\gamma\alpha^p.$$
By the admissible gradient transformations of the potential, the values $\tilde{c}^p$ and $\gamma$ can be set to zero. The solution has the form:
$$A^p=\frac{a^ph+\alpha^p\omega}{\Delta}, \quad A^\nu=0,$$
\begin{align}\label{40}
 & (g^{ij})=\begin{pmatrix}\frac{(a^0)^2+\varepsilon(\alpha^0)^2}{\Delta} & \frac{a^0a^1+\varepsilon\alpha^0\alpha^1}{\Delta} & 0 & 0\\\frac{a^0a^1+
  \varepsilon\alpha^0\alpha^1}{\Delta}
  & \frac{(a^1)^2+\varepsilon(\alpha^1)^2}{\Delta} & 0 & 0\\
   0 & 0 & \frac{\varepsilon_2}{\Delta} & 0 \\ 0 & 0 & 0 & \frac{\varepsilon_3}{\Delta}\end{pmatrix}.\end{align}
   This result was first obtained by Carter \cite{38}.

   $\mathbf{b})$ Matrix $\hat{G}_3$.

   We substitute the matrix $\hat{G}_3$ into the equation \eqref{23}. After some transformations in the equation \eqref{23} we get: $\quad h^1+\omega^1=\gamma(\alpha^{01}+a^{01}).\quad $
Hence, \quad $\gamma=\tilde{\gamma}$.\quad By the admissible gradient transformation of the potential $\quad \tilde{\gamma}\quad $ can be set to zero.  The final solution is:
 \begin{equation}\label{42} A^{0}=\frac{h+\omega}{\Delta}\quad A^1=A^{\nu}=0, \quad g^{ij}=\begin{pmatrix} \frac{a_0+\alpha_0}{\Delta} & \frac{a+\alpha}{\Delta} & 0 & 0 \\ \frac{a+\alpha}{\Delta} & 0  & 0 & 0 \\0 & 0 & \frac{\varepsilon_2}{\Delta} & 0 \\ 0 & 0 & 0 & \frac{\varepsilon_3}{\Delta} \end{pmatrix} . \end{equation}

 $\mathbf{c})$ Matrix $\hat{G}_5$. Substitute it in the equation \eqref{23}.  Denote: $ H_{p}=h^{p}+\sigma^p$.  After the reduction, we get:
 $$H_1=\sigma^1 \to h^1 = 0. $$ The equation \eqref{24} can be reduced to the form:
 \begin{align}\label{43}(H_0-\alpha^0\sigma)^2=(\rho+p)A, \quad (A=a+\alpha).\end{align}
 Equation \eqref{43} has a unique solution: \quad $H_0=\alpha^0\sigma+\hat{e}A.$ \quad
 By the admissible gradient transformation of the potential we vanish $\tilde{e}$. The solution has the form
$$A^0=\frac{a^0\omega}{\Delta}, \quad A^1= A^\nu=0.$$
   \section{Quadratic dependence between free functions.}

   To complete the classification, we have to consider the variant of a quadratic dependence between the functions $\omega^p$.  Using the same technique as before, from the equation \eqref{13} we find the relation connecting the functions $\omega^{p}$. Without limiting the generality, we assume that the free function is $\omega^0$. Then:
   \begin{align}\label{44} \omega^1=\gamma_0 + \gamma_{1}\omega^0 + \Sigma\quad (\Sigma^2 = \phi_2 {\omega^0}^2 + \phi_1 \omega^0 + \phi_0).\end{align}
   In the relations \eqref{44}, all functions except the free function $\omega^0$  are expressed in terms of $\alpha^{pq}$. If $\Sigma^2$ is a full square, we get the already considered version of the linear relationship between $\omega^{p}$. Therefore:

  \begin{align}\label{45}(\phi_1)^2-\phi_0 \phi_2 \ne 0.\end{align}
   Obviously, under this condition
   \begin{align}\label{46}\omega^0,\quad {\omega^0}^2,\quad \omega^0\Sigma,\quad \Sigma\end{align} -
     are linearly independent functions (with coefficients depending on $\alpha^{pq}$). These functions are included in the function $\omega^0$ as follows (see \eqref{13}):
   \begin{align}\label{47} \hat{\omega}=\rho+\tau_2{\omega^0}^2+2\tau_1\omega^0+\tau_0+2\xi_1\omega^0\Sigma+2\xi_0\Sigma.\end{align}
   Substituting \eqref{47} into the system \eqref{13} and equating the coefficients to the independent functions \eqref {47} to zero, we obtain the following systems of equations after some transformations:
\begin{equation}\label{48}
   \left\{\begin{array}{ll}
G^{00}\gamma_1-G^{01} =\xi_1 G,\quad\cr
G^{00}(\phi_2-{\gamma_1}^2)+G^{11}=(\tau_2-2\xi_1 \gamma_1)G;
\\\end{array}\ \right.\end{equation}

\begin{equation}\label{49}
   \left\{\begin{array}{ll}
G^{00}h^1-G^{01}h^{0}=\xi_0 G -G^{00}\gamma_0, \quad\cr
G^{01}h^{1}-G^{11}h^{0}=(-\tau_1+\gamma_1 \xi_0)G -G^{01}\gamma_0 + G^{00}\phi_1; \\\end{array}\ \right.\end{equation}

\begin{equation}\label{50}
   G^{00}((h^1+\gamma_{0})^2+\phi_0 )+G^{11}{h^0}^2-2G^{01}h^{0}(h^{1}+\gamma_0)=(\rho+p)G.\end{equation}
  The equations \eqref{48} are solved in the same way as in the case of linear dependence of the functions $\omega^p$. Using the classification \eqref{25}, the matrix $\hat{G}$ is found. A distinctive feature is that the completion of this stage does not mean, as before, automatically obtaining a particular case of the desired potential, since the system \eqref{49} and the equation \eqref{50} non-trivially include the functions defining $\omega^1$. Now, to find the potential, it is necessary to satisfy all the remaining equations. To do this, the found matrices $\hat{G}_a$ are substituted in them and a solution is sought. The second distinguishing feature is that the system \eqref{49} immediately implies absence in the potential of a free function that depends only on  $u^3$ since from $\quad \det{\hat{G}} \ne 0 \to $ the functions  $ h^p$ are expressed in terms of $a^{pq}$. The third feature is that the obtained solutions must satisfy the condition \eqref{45}.

Omitting the obvious, but rather cumbersome calculations, we present the final solutions for the \eqref{48}-\eqref{50}:

\begin{align}\gamma_0 =\gamma_{1}=\phi_1=0, \quad \phi_0=-\varepsilon,\quad \phi_2=\tilde{a}.\end{align}
\begin{align}\label{51}(\omega^1)^2=\sqrt{\tilde{a}-\varepsilon(\omega^0)^2}.\end{align}
\begin{equation}\label{52}\hat{G}=\begin{pmatrix} a & 0 \\ 0 & \varepsilon a \end{pmatrix},\quad \varepsilon=+1,-1.\end{equation}
The metric tensor and electromagnetic potential have the form:
\begin{align}\label{53}\left\{\begin{array}{ll}(g^{ij})=\begin{pmatrix}\displaystyle\frac{a}{\Delta} & 0 & 0 & 0\\ 0 &\displaystyle\frac{\varepsilon a}{\Delta} & 0 & 0\\0 & 0 & \displaystyle\frac{\varepsilon_2}{\Delta} & 0 \\ 0 & 0 & 0 & \displaystyle\frac{\varepsilon_3}{\Delta}\end{pmatrix};\cr \mathbf{a}) \varepsilon=1,\quad A^0=\displaystyle\frac{\tilde{e}sin{\sigma}}{\Delta},\quad A^1=\displaystyle\frac{\tilde{e}cos{\sigma}}{\Delta}, \quad A^{\nu}=0;
\cr \mathbf{b}) \varepsilon=-1,\quad A^0=\displaystyle\frac{\tilde{e}sh{\sigma}}{\Delta},\quad A^1=\displaystyle\frac{\tilde{e}ch{\sigma}}{\Delta}, \quad A^{\nu}=0.\\\end{array}\ \right.\end{align}

\section{Discussion}
The main result of the classification is to obtain all metrics and potentials of the external electromagnetic field that have sufficient symmetry to perform an exact integration of the Hamilton-Jacobi equation for a charged test particle. Currently, a large number of exact solutions of Einstein's equations are known \cite{41}. However, only a small part of them have such property, and they are the main object of research. The additional symmetry of Stackel spaces, which allows us to separate variables in the Hamilton-Jacobi equation for a charged test particle, can be  applied for integrating the vacuum equations of the gravitational field both in General relativity (for example, when studying the problem of axion fields \cite{42}) and in alternative theories. For convenience of use metrics: \quad $(ds)^2=g_{ij}du^i du^j,$ \quad covariant components of electromagnetic potentials: $A_i$, and separated systems whose solutions determine the complete integrals of the equation \eqref{1}
 \begin{align}\label{a11}\hat{H}_2 = \lambda_3\phi + \lambda_2, \quad \hat{H}_3 = \lambda_3f-\lambda_2.\end{align}
 are given below.

    The separated systems  have the form:
   \begin{align}\label{a11}\hat{H}_2 = \lambda_3\phi + \lambda_2, \quad \hat{H}_3 = \lambda_3f-\lambda_2.\end{align}
     In conclusion, we would like to give a complete summary of the results. For each solution of the equation \eqref{13} we give the metric $(ds)^2=g_{ij}du^i du^j $, the covariant components of the electromagnetic potential \quad $A_p$ and the function $ \hat{H}_\nu $, which define the integrals of motion quadratic in momenta  in accordance with :   \begin{align}\label{54}\hat{X}_2 =\frac{fH_2-\phi H_3}{\Delta} \quad,\end{align} where
 $$ \hat{H}_2 =\alpha^{pq}p_{p}p_{q}+\varepsilon_2 p_2^2+2\omega^q p_q + \hat{\omega},\quad \hat{H}_3 =h^{pq}p_{p}p_{q}+\varepsilon_3 p_3^2+2h^q p_q + \hat{h}.$$
 Recall that $\quad  A_2=A_3=0, \quad \Delta = \phi+ f $.

 $\mathbf{I}$.
\begin{equation}\label{55}
   \left\{ \begin{array}{ll}
(ds)^2=(\displaystyle\frac {({\alpha^1}^2 + \varepsilon{a^1}^2 ){du^0}^2 + ({\alpha^0}^2 + \varepsilon{a^0}^2) {du^1}^2 \vspace{3mm}
-2(\alpha^0\alpha^1 +\varepsilon a^0 a^1)du^0du^1} {(\alpha^1a^0-a^1\alpha^0)^2}+&
\cr \varepsilon_2{du^2}^2 + \varepsilon_3{du^3}^2)\Delta.\quad \cr\vspace{1mm}
 A_0 = \displaystyle\frac{\varepsilon h\alpha^1-\omega a^1}{(\alpha^1 a^0-a^1 \alpha^0)}, \quad\cr \vspace{2mm}
A_1 = \displaystyle\frac{-\varepsilon h\alpha^0+\omega\alpha^0}{(\alpha^1a^0-a^1\alpha^0)}. \quad \cr \vspace{5mm}
 \hat{H}_2=\varepsilon_2{p_2}^2 + \alpha^{p}\alpha^{q}p_pp_q+2\alpha^{q}p_q\omega + \varepsilon\omega^2 \quad \cr
 \hat{H}_3=\varepsilon_3{p_3}^2 + \varepsilon a^{p}a^{q}p_pp_q+2a^{q}p_q h + h^2.\\\end{array}\ \right.\end{equation}
$\mathbf{II}$.
\begin{equation}\label{56}
   \left\{ \begin{array}{ll}
(ds)^2=(\displaystyle\frac {2(\alpha + a)du^0du^1-(\alpha_0 + a_0){du^1}^2} {(\alpha+a)}+&
\cr \varepsilon_2{du^2}^2 + \varepsilon_3{du^3}^2)\Delta.\quad \cr\vspace{1mm}
 A_0 = 0, \qquad A_1 = \displaystyle\frac{ h+\omega}{(\alpha+a)}.\cr \vspace{5mm}
 \hat{H}_2=\varepsilon_2{p_2}^2 + \alpha_0 p_0^2+2\alpha p_0 p_1 + 2\omega p_0, \quad \cr
  \hat{H}_3=\varepsilon_3{p_3}^2 + a_0 p_0^2+2a p_0 p_1 + 2h p_0  \quad \cr
 \\\end{array}\ \right.\end{equation}
 $\mathbf{III}$.
\begin{equation}\label{57}
   \left\{ \begin{array}{ll}
(ds)^2=(\displaystyle\frac{(\alpha + a){du^1}^2 + \varepsilon({\alpha^0} {du^1} - {\alpha^1} {du^0})^2}{(\alpha+a){\alpha^1}^2}+&
\cr \varepsilon_2{du^2}^2 + \varepsilon_3{du^3}^2)\Delta.\quad \cr\vspace{1mm}
 A_0 = 0, \qquad A_1 = \displaystyle\frac{\omega}{\alpha^1}.\cr \vspace{3mm}
 \hat{H}_2=\varepsilon_2{p_2}^2 + \alpha p_0^2+(\alpha^q p_q+\omega)^2 {du^0})^2 \quad
  \hat{H}_3=\varepsilon_3{p_3}^2 + a p_0^2 \quad \cr
 \\\end{array}\ \right.\end{equation}

 $\mathbf{IV}$.
\begin{equation}\label{58}
   \left\{ \begin{array}{ll}
(ds)^2=(\displaystyle\frac{\alpha^{11} {du^0}^2 +\alpha^{00} {du^1}^2 -2\alpha^{01} du^{0}du^{1}}{(\alpha^{00}\alpha^{11}-{\alpha^{01}}^2)}+&
\cr \varepsilon_2{du^2}^2 + \varepsilon_3{du^3}^2)\Delta.\quad \cr\vspace{1mm}
 A_{0} = \displaystyle \frac{\omega^{0} \alpha^{11}-\omega^{1} \alpha^{01}}{(\alpha^{00} \alpha^{11}-{\alpha^{01}}^2)}.\cr \vspace{5mm}
 A_{1} =
 \displaystyle
  \frac{\omega^{1} \alpha^{00}-\omega^{0} \alpha^{01}}{(\alpha^{00} \alpha^{11}-{\alpha^{01}}^2)}.\cr \vspace{5mm}
\tilde{H}_2=\varepsilon_2 {p_2}^2+\alpha^{pq}{p_p}{p_q} +2\omega^p p_p+\cr \vspace{1mm}
\frac{\alpha^{00}(\omega^1)^2+\alpha^{11}(\omega^0)^2-2\alpha^{01}\omega^0\omega^1}{(\alpha^{00} \alpha^{11}-{\alpha^{01}}^2)}.\quad
\hat{H}_3=\varepsilon_3{p_3}^2 \quad \cr \vspace{1mm}
 \\\end{array}\ \right.\end{equation}
 $\mathbf{V}$.
\begin{equation}\label{59}
   \left\{ \begin{array}{ll}
(ds)^2=(\displaystyle\frac{{du^0}^2 +\varepsilon{du^1}^2 }{a}+\varepsilon_2{du^2}^2 + \varepsilon_3{du^3}^2)\Delta.\quad \cr\vspace{3mm}
\mathbf{a}) \quad \varepsilon=-1,\quad  A_{0} = \displaystyle  \frac{sh{\omega}} {a},\quad A_{1}=\frac{ch{\omega}}{a} .\cr \vspace{5mm}
\displaystyle\tilde{H}_2=\varepsilon_2 {p_2}^2+2(p_{0}sh{\omega}-p_{1}ch{\omega}), \quad
\displaystyle \tilde{H}_3=\varepsilon {p_3}^2+a({p_0}^2-{p_1}^2)+\frac{1}{a^2} ,\cr \vspace{1mm}
\mathbf{b}) \quad \varepsilon=1,\quad  A_{0} = \displaystyle \frac{sin{\omega}}{a},\quad A_{1}=\frac{cos{\omega}}{a} .\cr \vspace{5mm}
\tilde{H}_2=\varepsilon_2 {p_2}^2+2(p_{0}sin{\omega}+p_{1}cos{\omega}), \quad
\tilde{H}_3=\varepsilon {p_3}^2+a({p_0}^2+{p_1}^2)+\frac{1}{a^2} ,\cr

 \\\end{array}\ \right.\end{equation}

\section{Conclusions}

 Thus, all space-time metrics and electromagnetic potentials that allow complete separation of variables of type (2.0) in the Hamilton-Jacobi equation \eqref{1} for a charged test particle moving in an external electromagnetic field are found. The complete sets of mutually commuting vector and Killing tensor fields and the complete sets of motion integrals are defined.  Note that the same problem has been solved for the Stackel spaces of type (1.0) (\cite{43}) and for the Stackel spaces of type (2.1) (\cite{44}).

acknowledgments {This work was supported by the Ministry of Science and Higher Education of the Russian Federation, project FEWF-2020-0003.}




\end{document}